\documentclass[10pt]{iopart}
\usepackage{graphicx}
\usepackage{ulem}
\usepackage{xcolor}
\usepackage{soul}
\usepackage[numbers,sort&compress]{natbib}
\usepackage{float}
\usepackage[utf8]{inputenc}
\usepackage[T1]{fontenc}
\usepackage{array}
\usepackage{dcolumn}% Align table columns on decimal point
\usepackage{bm}% bold math
\usepackage{siunitx}
\usepackage{amssymb,latexsym}
\usepackage{afterpage}
\usepackage{rotating}
\usepackage[switch]{lineno}

\begin{document}

\title[]{Symmetry Guided Band-Gap Opening via Periodic Topological Defects in Graphene\\
}
%=====================================================
\author{
D. N. Garzon$^{1,2,3 \ast, \ddag,\dag}$, 
Leonel Cabrera-Loor$^{1,4, \ast, \S,\dag}$,
Jacopo Gliozzi$^{3}$, 
Marco Fronzi$^{4}$, 
Catherine Stampfl$^{4}$ and
Henry P. Pinto$^{1,\dag}$
}
\address{$^1$ CompNano Group, School of Physical Sciences and Nanotechnology, Yachay Tech University, 100119-Urcuqui, Ecuador}
\address{$^2$ Illinois Center for Advanced Studies of the Universe, University of Illinois Urbana-Champaign, Urbana, Illinois 61801, USA}
\address{$^3$The Grainger College of Engineering,
Department of Physics, University of Illinois Urbana-Champaign, Urbana, Illinois 61801, USA}
\address{$^4$ School of Physics, The University of Sydney, Camperdown, NSW 2006, Australia}
%=====================================================

\vspace{10pt}
\begin{indented}
\item[]$^\ast$ These authors contributed equally to this work.
\item[]$^\ddag$ Present address: Department of Physics, University of Illinois at Urbana-Champaign, Urbana, IL 61801, USA.
\item[]$\S$ Present address: School of Physics, The University of Sydney, Camperdown, NSW 2006, Australia.
\item[]$\dag$ Corresponding authors.
\end{indented}

\vspace{10pt}
\begin{indented}
%\item[]August 2017 (minor update March 2024)
\item[] \textbf{Keywords:} topological defects, patterned graphene, Stone--Wales defect, flower--like defect, band-gap, density functional theory (DFT).
\end{indented}

\begin{abstract}
Graphene lacks an intrinsic band-gap, which limits its use in electronic applications. Here we demonstrate that periodic arrays of topological defects can open and control a band-gap in a predictable manner governed by defect 
spacing and lattice symmetry. Using first-principles density functional theory calculations supported by tight-binding models, we investigate graphene superlattices containing Stone--Wales and flower--like defects over a range of $N \times N$ periodicities, where $N$ determines the defect separation. We show that band-gap opening occurs only when translation symmetry is reduced in a specific way: for supercells with $N$ a multiple of three, Brillouin-zone folding brings the Dirac cones at $K$ and $K'$ to the same momentum in the reduced Brillouin zone. In particular, flower-like defect superlattices produce larger and tunable band-gaps, whose magnitude decreases systematically with increasing defect separation and approaches zero in the dilute-defect limit. These results establish a predictive framework for band-gap engineering in defect-patterned graphene and clarify the microscopic mechanism underlying gap formation in
periodically reconstructed lattices.
\end{abstract}

\vspace{2pc}
\ioptwocol
%
%\linenumbers 
\section{ Introduction}\label{sec:introduction}
Because of its remarkable electronic, mechanical, and structural properties, graphene has attracted significant attention for a broad range of technological applications, including flexible electronics, sensors, and energy devices \cite{Novoselov666,geim2007rise,neto2009electronic,bonaccorso2010graphene,lee2009elastic}. However, the absence of an intrinsic band-gap poses a major limitation for its use in conventional nanoelectronic and digital logic devices, which require clear on/off switching behavior \cite{Dvorak2013}. Therefore, band-gap engineering remains a critical challenge for enabling graphene’s integration into semiconductor technology \cite{graphene_band_gap_problem}. 
Structural irregularities in two-dimensional (2D) materials can strongly affect their electronic properties, including a band-gap opening~\cite{banhart2011structural}. One promising strategy to induce a finite band-gap involves the controlled introduction of topological defects, which can modify the electronic structure \cite{amara2007scanning, guinea2010gap, park2015band, zhang2023atomic}. These defects arise from the local rearrangement of the hexagonal lattice into non-hexagonal rings (e.g., pentagons and heptagons), leading to structural distortions that can significantly alter the electronic band structure and the density of states (DOS) \cite{Cockayne, Cockayne2012, rasool2011atomic}.

Previous studies suggest that extended topological defects, sometimes described as \textit{graphene superlattices}, may offer a route to engineer graphene-based nanostructures with tailored conductivity~\cite{Pinto2020}. Building on this idea, several strategies have been explored to induce semiconducting behavior in graphene, including the periodic patterning of topological defects~\cite{shirodkar2012electronic,yan2013electronic,Dvorak2013,kim2010fabrication}. Nevertheless, it remains unclear why some periodic topological defect superlattices develop a band-gap while others remain metallic.

A notable class of topological defects in graphene are grain boundary loops (GBLs), which can be understood as rotationally reconstructed regions formed by cutting out a portion of the graphene lattice and rotating it relative to the pristine lattice~\cite{Cockayne}. Among these, two well-studied examples are the Stone–Wales defect (SWD) and the flower-like defect (FLD). The SWD represents the smallest possible rotational defect and has been shown to modify key properties of graphene, including its electronic behavior~\cite{boukhvalov2008chemical,kang2008effect}. The FLD, meanwhile, is predicted to have the lowest formation energy per dislocation core of any known topological defect in graphene, suggesting favorable energetic stability under suitable conditions. Experimental evidence for FLDs has been reported through scanning tunneling microscopy (STM), revealing characteristic flower-like patterns \cite{Cockayne}.

The electronic properties of graphene superlattices, including the existence of a band-gap, are governed by the presence or absence of symmetries. In pristine graphene, the gapless Dirac points are protected by a combination of time-reversal, inversion, and translation symmetry~\cite{RMP_2009}. Opening a gap therefore requires the breaking of one or more of these symmetries.
Superlattices of topological defects automatically break translation symmetry, and can also break inversion symmetry depending on their pattern. 
Nevertheless, a predictive framework that connects the symmetries of the defect superlattices to their electronic structure, especially for FLDs, is missing.

In this work, building on the preliminary results of Garzon \citep{garzon2020}, we investigate the atomic and electronic structure of \textit{graphene superlattices} formed by periodic arrangements of SWD and FLD. Our aim is to understand how the introduction of these topological defects modifies the symmetry and electronic properties of graphene, and how these changes can be exploited for band-gap engineering.

We perform first-principles density functional theory (DFT) calculations, as implemented in the Vienna \textit{Ab initio} Simulation Package (VASP) \cite{vasp,vasp2-kresse1994norm}, using the regularized-restored Strongly Constrained and Appropriately Normed (r$^2$SCAN) meta-generalized gradient approximation (meta-GGA) exchange-correlation functional \cite{furness2020accurate} under periodic boundary conditions. We additionally use the Heyd--Scuseria--Ernzerhof 2006 (HSE06) hybrid functional \cite{heyd2003hybrid,Krukau2006} for the structures exhibiting a band-gap. We evaluate the thermodynamic stability of the defective superlattices through total- and formation-energy analyses, and investigate their electronic structure and local electronic signatures by means of symmetry analysis, projected density of states (PDOS), and simulated scanning tunneling microscopy (STM) images. This combined approach allows us to identify the structural and electronic mechanisms responsible for band-gap opening in defective graphene superlattices. We also employ Density functional tight-binding (DFTB+) calculations \cite{dftb+} to investigate the band-gap-opening rule in larger FLD superlattices, thereby extending the analysis beyond the computational limits of DFT with the r$^2$SCAN functional. 

In addition to the DFT calculations, we develop a tight-binding (TB) model to provide a theoretical framework linking defect geometry and symmetry breaking to electronic band-gap formation. Based on this analysis, we propose a simple predictive rule for the band-gap as a function of defect spacing and validate it through direct comparison with DFT results. Our work therefore addresses a long-standing problem of how to properly pattern graphene to induce band-gaps, pointing to practical routes for symmetry-guided design of graphene with controllable and desirable electronic band-gap openings. 

Throughout this work, when referring to the periodic arrays of topological defects in graphene, we use the term  \textit{superlattices}. The remainder of this paper is organized as follows. In Sec.~\ref{sec:methods}, we describe the construction of the SWD and FLD superlattices and summarize the computational methods used in this work. In Sec.~\ref{sec:theory}, we introduce the tight-binding model and derive the band-gap-opening rule. In Sec.~\ref{sec:results}, we present and discuss the structural, electronic, and symmetry results, including the mechanism responsible for gap opening. Finally, in Sec.~\ref{sec:conclusions}, we summarize our main conclusions. 

\begin{figure*}[ht]
\centering
\includegraphics[width=0.9\linewidth]{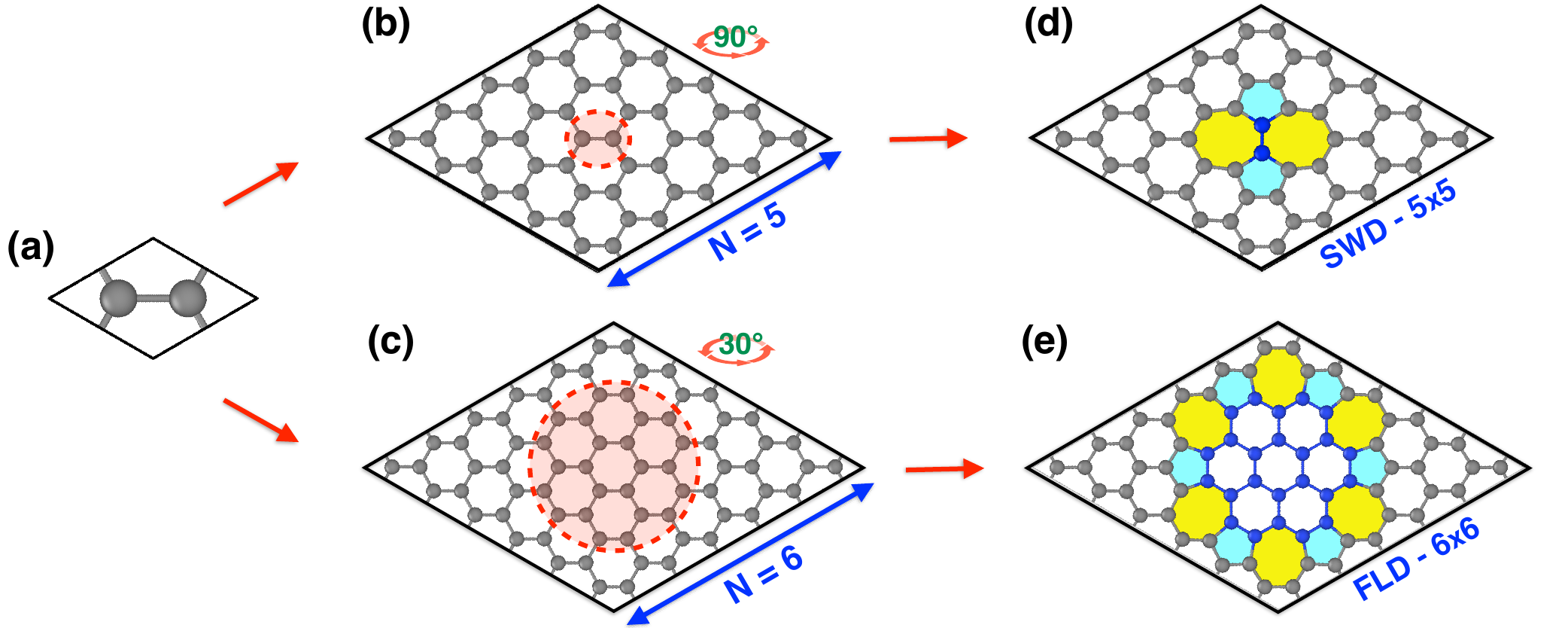}
\caption{Schematic illustration of the workflow to construct the graphene defective superlattices considered in this work. The workflow starts from (a) the primitive graphene cell, from which an $N\times N$ supercell is generated, as illustrated for (b) $N=5$ and (c) $N=6$. A defect region is then defined at the supercell center (red shaded area with dashed outline), involving 2 atoms for the SWD and 24 atoms for the FLD. The corresponding defect is created by rotating the selected atoms by 90$^\circ$ (SWD) and 30$^\circ$ (FLD), yielding (d) SWD–5$\times$5 and (e) FLD–6$\times$6. Carbon atoms involved in the rotation are shown in blue, and the resulting non-hexagonal rings are highlighted to emphasize the defect topology, with heptagons in yellow and pentagons in sky blue.}
\label{fig:defect_workflow}
\end{figure*}

\section{Computational Methods}\label{sec:methods}

In this section, we outline the methodology used to construct and analyze the defective graphene superlattices. 

We first describe the geometric construction of the SWD and FLD and the generation of their periodic superlattices. We then detail the lattice-optimization procedure and the first-principles computational settings employed in VASP, followed by the extended calculations performed with DFTB+ for larger systems. Finally, we introduce the simplified tight-binding model used to provide a theoretical interpretation of the band-gap behavior.

\subsection{Defect structure}\label{subsec:defect_structure}

To generate the SWD structure, an in-plane rotation of $\ang{90}$ is applied to two carbon atoms about the midpoint of their shared bond. This bond rotation reconstructs the local hexagonal network, transforming four adjacent hexagons into a pair of pentagons and a pair of heptagons, as shown in Fig.~\ref{fig:defect_workflow}d. For the FLD, a similar procedure is followed; however, in this case, 24 carbon atoms within the graphene lattice participate in the reconstruction, and their hexagonal arrangement is rotated by $\ang{30}$. The bonds are subsequently reconnected, preserving the total number of carbon atoms while generating a central cluster composed of alternating pentagons and heptagons embedded within the surrounding hexagonal lattice (see Fig.~\ref{fig:defect_workflow}e). Interestingly, the FLD retains $D_{6h}$ symmetry, with no unsatisfied bonds, and each carbon atom remains threefold coordinated.

\subsection{Superlattices}
\label{subsec:superlattices}

We constructed several monolayer superlattices containing periodic arrangements of the SWD and FLD. We generated $N$$\times$$N$ graphene supercells by replicating the pristine primitive cell $N$ times along each in-plane direction (see Fig.~\ref{fig:defect_workflow}). Within each supercell, a single defect was introduced, yielding a periodic defective superlattice in which the defect separation is determined by $N$. We examined a range of periodicities, considering FLD supercells with $N \in \{5,6,7,8,9\}$ and SWD supercells with $N \in \{3,4,5,6\}$, which includes the smallest defect-compatible cells and a systematic set of larger superlattices.

\subsection{Lattice Optimization}
\label{subsec:lat_opt}

To obtain the equilibrium geometry of each superlattice, the in-plane lattice parameters were optimized by fitting a 2D equation of state (EOS) to total-energy data~\cite{EOS_andrew2012mechanical}. This procedure was applied independently to each superlattice in order to ensure a consistent determination of the equilibrium lattice constant for subsequent structural and electronic analyses.

First, a series of structures was generated by systematically varying the in-plane lattice constants around an initial estimate, keeping the vacuum spacing along the $z$-direction fixed (15 \AA). For each choice of in-plane lattice variation of the superlattice, the lattice vectors were held fixed and only the internal atomic coordinates were relaxed, so that the total energy reflected the response to the imposed in-plane strain without additional lattice changes. The converged total energies from this set of configurations were then fitted to the 2D EOS, from which the equilibrium in-plane lattice parameters were extracted.

Finally, using the fitted equilibrium lattice constants, a final ionic relaxation (with fixed cell vectors) was performed to obtain the minimum-energy atomic structure. This workflow improves the robustness of the optimized geometries and avoids relying on a single direct relaxation from an initial lattice guess.

\subsection{VASP details}

This subsection summarizes the computational settings used for the DFT calculations performed with the Vienna \textit{Ab initio} Simulation Package (VASP) \cite{vasp,vasp2-kresse1994norm} for the pristine and defective graphene superlattices.

To obtain a reliable description of the defective graphene superlattices, we employed the meta-GGA $r^2$SCAN functional \cite{furness2020accurate}. This choice offers two main advantages. First, within the generalized Kohn--Sham framework, meta-GGAs can mitigate self-interaction and delocalization errors, often improving the separation between occupied and unoccupied states and reducing systematic errors in relative energies compared with the local density approximation (LDA) and standard generalized gradient approximation (GGA) functionals such as Perdew--Burke--Ernzerhof (PBE) \cite{kingsbury2022flexible,zhang2025advances}. Second, $r^2$SCAN retains the high accuracy of the original SCAN functional across diverse bonding environments while significantly improving numerical stability through regularization and restoration of exact constraints \cite{scan2015,bartok2019regularized,furness2020accurate}.

This refinement ensures robust convergence and maintains a tractable computational cost, making it well suited for the demanding requirements of large-supercell simulations. The core–valence interactions were described using the projector augmented-wave (PAW) method. We adopted a convergence criterion of 1~meV~atom$^{-1}$ for both the plane-wave kinetic-energy cutoff and the $k$-point sampling. Accordingly, a cutoff of 950~eV was selected to ensure high numerical precision (Fig.~\ref{fig:cutoff}). For the Brillouin zone integration, Monkhorst--Pack meshes \cite{monkhorst1976special} with a reciprocal-space $k$-point spacing of $0.022\times 2\pi$~\AA$^{-1}$ were adopted for all structures (Fig.~\ref{fig:kpts}). A vacuum spacing of 15~\AA\ was introduced along the out-of-plane ($z$) direction to suppress spurious interactions between periodic images. Electronic self-consistency was converged to $10^{-6}$ eV, and ionic relaxations were performed until residual forces on all atoms were below 0.01 eV \AA$^{-1}$.

Structural relaxations were performed without enforcing any symmetry constraints. %(\texttt{ISYM=0}). 
The crystallographic symmetry of pristine graphene and each defective graphene superlattice was then determined from the fully relaxed atomic geometries using two independent symmetry-detection approaches: the symmetry-finding routine implemented in \textsc{VASPKIT} \cite{wang2021vaspkit} and the \textsc{FINDSYM} software \cite{stokes2005findsym}. To avoid spurious symmetry lowering caused by residual numerical displacements in large defective supercells, a symmetry tolerance of $10^{-2}$ \AA\ was adopted. The reported space group and point group therefore correspond to the relaxed geometries within this tolerance. Both approaches yielded the same space-group assignments for all structures considered. 

In addition, simulated STM images of the  FLD superlattices  were generated from the VASP electronic-structure outputs (including the wavefunctions) using the Tersoff--Hamann approximation, as implemented in the bSKAN software~\cite{bskan}.

\subsection{DFTB+ details}
To further extend the study to larger superlattices and to corroborate important trends obtained from the \textit{ab initio} calculations, we performed additional calculations using DFTB+ (version 22.2) \cite{dftb+} for the FLD superlattices. First, DFTB+ was used to benchmark the band-gap trends obtained with VASP for FLD supercells with $N \in \{5,6,7,8,9\}$. After this consistency check, the FLD size range was extended by examining larger superlattices, from 10$\times$10 to $15\times15$, using DFTB+.

In the present work, we employed the GFN2-xTB extended tight-binding Hamiltonian \cite{xTBH,GFN2xTB}. 
Structural relaxations of the superlattices were carried out using a self-consistent charge (SCC) convergence tolerance of $10^{-6}$. For all DFTB+ calculations, a $k$-point spacing of $0.019\times 2\pi$~\AA$^{-1}$ was used.

\section{ Tight-binding model}\label{sec:theory}

In addition to the detailed DFT band structure calculations, we also employed a simplified tight-binding model to understand the origin of the band-gaps in certain defect superlattices. For both FLD and SWD structures, we started with a standard tight-binding model of pristine graphene: uniform nearest-neighbor hopping on a hexagonal lattice~\cite{RMP_2009}. Depending on the defect type, we then rotated the relevant carbon atoms and reconnected the bonds as detailed in Sec.~\ref{subsec:defect_structure}. The resulting lattice retains only nearest-neighbor bonds, as in pristine graphene, and differs from the pristine lattice through the local connectivity and hopping amplitudes near the defect cores.

Starting from this tight-binding model, we allowed the hopping amplitudes on the rotated bonds at the defect core to differ slightly from the pristine value. This was done both to capture the effects of bond relaxation near the defects and for greater generality. Specifically, we considered hopping strengths $t_\text{defect} \in [0.8 t_\text{pristine}, 1.2 t_\text{pristine}]$ for the rotated bonds at the defect core.

Using this simplified tight-binding model, we recovered the same pattern of band-gap opening and closing observed in the DFT results (Sec.~\ref{subsec:electronic}). Moreover, the model yielded approximate locations of Dirac points in SWD lattices, where they can shift away from high-symmetry 
points~\cite{shirodkar2012electronic}, facilitating the calculation of the band-gap with DFT. Lastly, the model provided a basis for the theoretical argument for band-gap opening that follows.

\subsection{Band-gap opening rule}

Based on the perturbative tight-binding model proposed by Dvorak \textit{et al.}~\cite{Dvorak2013}, we analyze the conditions under which periodic arrays of grain boundary loops can open a band-gap. Within this framework, the effects of the grain boundary loops can be modeled as a periodic potential. The periodicity of this superlattice is set by two vectors, $\bm{R}_1 = n_1 \bm{a}_1 + m_1 \bm{a}_2$ and $\bm{R}_2 = n_2 \bm{a}_1 + m_2 \bm{a}_2$, where $n_{1}, m_{1}, n_{2}$, and $m_{2}$ are integers and $\bm{a}_1$ and $\bm{a}_2$ are the lattice vectors of pristine graphene.
% then as a general rule, 
Following Ref.~\cite{Dvorak2013}, patterned lattices can open a band-gap when
\begin{equation}
\left\{
\begin{array}{l}
(n_{1}+m_{1}) \pm \frac{1}{3}(n_{1}-m_{1}) = 2\alpha \\\\
(n_{2}+m_{2}) \pm \frac{1}{3}(n_{2}-m_{2}) = 2\beta
\end{array}
\right.
\end{equation}
where $\alpha$ and $\beta$ are integers.
These conditions are satisfied only if
\begin{equation}
\label{eq:ruledvorak}
  n_{1}-m_{1} = 3p ,
  n_{2}-m_{2} = 3q,
\end{equation}
where $p$ and $q$ are integers. 

The superlattices containing the grain boundary loops are defined with lattice vectors  in a zigzag direction. As a result, they are characterized by a pair of integers ($Z_1, Z_2$), with ($n_1, m_1, n_2, m_2$)=($Z_1, 0, 0, Z_2$). We can then rewrite Eq.~\ref{eq:ruledvorak} as 
\begin{equation}
\label{eq:ruleFLD}
  n_{1} = 3p,
  m_{2} = 3q,
\end{equation}
For the examples we consider, this implies that $N$$\times$$N$ superlattices can generate a band-gap when $N$ is a multiple of three.

%=================================================
\begin{figure}[htbp]
\centering
\includegraphics[width=0.9\linewidth]{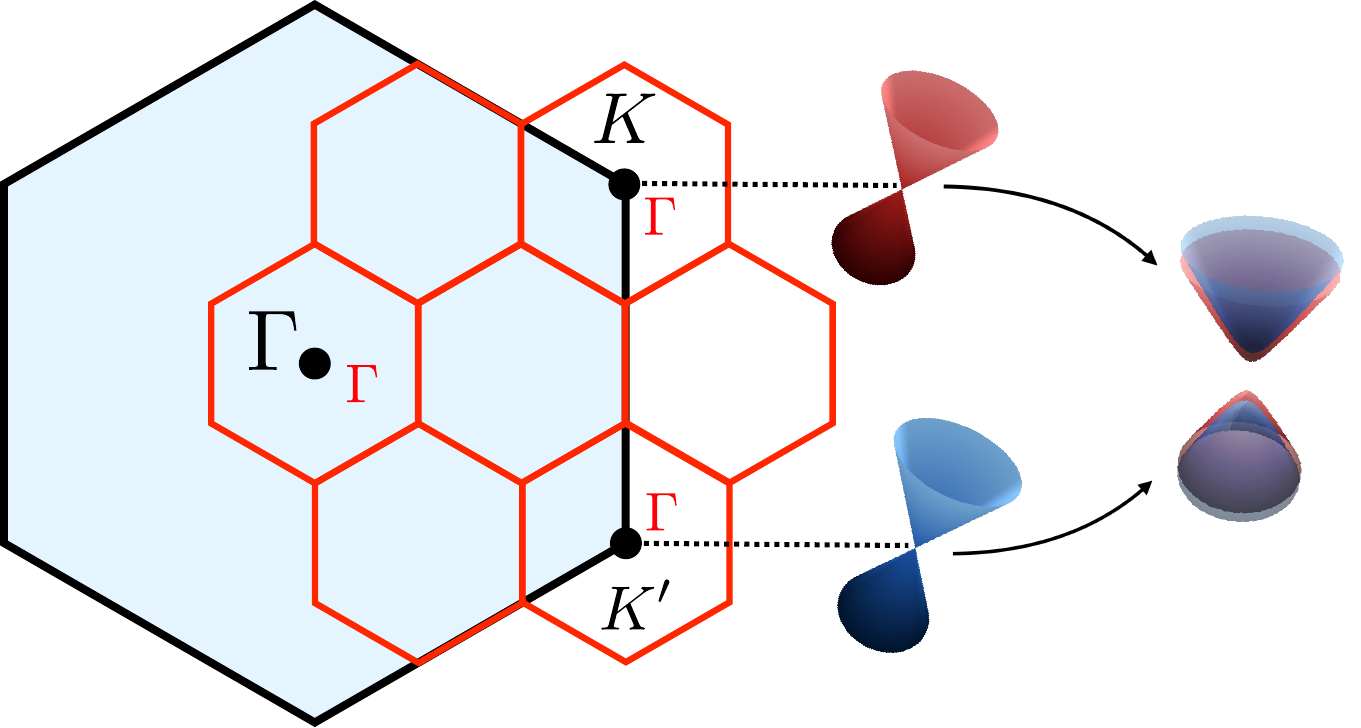}
\caption{\label{fig:bzfold} 
Schematic illustration for gap-opening mechanism in $3N \times 3N$ defective superlattices. As the unit cell is tripled, the Brillouin zone is correspondingly folded by a factor of three. The $K$ and $K'$ points of pristine graphene (black) both map to $\Gamma$ in the folded Brillouin zone (red), allowing the two distinct Dirac cones to couple and open a gap.}
\end{figure}
%=================================================

The gap opening condition in Eq.~\ref{eq:ruleFLD} can also be restated in terms of the Dirac points of pristine graphene. Without any defects, the two inequivalent Dirac points are found at $K$ and $K'$ in the Brillouin zone. When defects are added, the periodicity of the unit cell is increased and the corresponding Brillouin zone is folded. If the superlattice is such that $K$ and $K'$ fold to the same $k$-point in the new Brillouin zone, the two Dirac points can couple and gap out~\cite{guinea2010gap}, as also visually depicted in Fig.~\ref{fig:bzfold}. 

Concretely, the locations of the Dirac points can be written in terms of the reciprocal lattice vectors of pristine graphene, $\bm{b}_1$ and $\bm{b}_2$:
\begin{equation}
\bm{K} - \bm{K}' = -\frac{1}{3}\bm{b}_1 + \frac{1}{3}\bm{b}_2 .
\end{equation}

If the unit cell is tripled in both directions by the superlattice, then $\bm{b}_1, \bm{b}_2 \rightarrow \bm{b}_1/3, \bm{b}_2/3$, and the difference between $K$ and $K'$ becomes a reciprocal lattice vector. Consequently, $K$ and $K'$ map to the same point in the reduced Brillouin zone, and a generic coupling between the two cones can open a band-gap.

\section{Results and Discussion}\label{sec:results}

In this section,  we report the lattice parameters, cohesive energies, band structures, and density of states (DOS) of these defects in various superlattice configurations. We then analyze the influence of symmetry on the electronic structure and its relation to the band-gap opening in these lattices.

\subsection{Structural properties}\label{subsec:structural}

The optimized structures exhibit an in-plane lattice expansion for both FLD and SWD superlattices relative to pristine graphene, as reflected in the optimized lattice parameter ($a_{\mathrm{op}}$) reported in Tab.~\ref{tab:table-general}. 

%=================================================================
\begin{table*}[ht]
\centering
\caption{\label{tab:table-general}
Computed structural parameters and electronic gaps for pristine graphene and the defective superlattices. Unless otherwise indicated, the values were obtained with $r^2$SCAN+rVV10.  Here, $a_{\mathrm{op}}$ denotes the optimized lattice constant, $N_{\mathrm{atom}}$ the number of atoms, Defect A.F. the defect atomic fraction (\%), $E_{c}$ the cohesive energy per atom, and $E_{g}$ the electronic band-gap. The space and point groups were identified using a symmetry precision of $10^{-2}$ \AA.}
\footnotesize
\begin{tabular}{@{}cccccccc}
\br
System & $a_{\mathrm{op}}$ (\AA) & $N_{\mathrm{atom}}$ & Defect A.F. (\%) & Space Group & Point Group & $E_{c}$ (eV) & $E_{g}$ (eV) \\ \mr
SWD–3$\times$3 & 7.464  & 18 & 11.11 & $Cmmm$ (65) & $D_{2h}$ & 7.19 & 0.30 (0.62)$^{a}$ \\
SWD–4$\times$4 & 9.892  & 32 & 6.25 & $Cmmm$ (65) & $D_{2h}$ & 7.40 & 0.00$^{*}$ \\
SWD–5$\times$5 & 12.334 & 50 & 4.00 & $Cmmm$ (65) & $D_{2h}$ & 7.48 & 0.00$^{*}$ \\
SWD–6$\times$6 & 14.782 & 72 & 2.78 & $Cmmm$ (65) & $D_{2h}$ & 7.52 & 0.001 (0.06)$^{a}$ \\ \mr
FLD–5$\times$5& 12.395 & 50 & 48.00 & $P6/mmm$ (191) & $D_{6h}$ & 7.43 & 0.00 \\ 
FLD–6$\times$6 & 14.817 & 72 & 33.33 & $P6mm$ (183)   & $C_{6v}$ & 7.49 & 0.62 (0.85)$^{a}$ \\ 
FLD–7$\times$7& 17.278 & 98 & 24.49& $P6/mmm$ (191) & $D_{6h}$ & 7.52 & 0.00 \\
FLD–8$\times$8& 19.725 & 128 & 18.75 & $P6/mmm$ (191) & $D_{6h}$ & 7.54 & 0.00 \\
FLD–9$\times$9& 22.173 & 162 & 14.81 & $P6/mmm$ (191) & $D_{6h}$ & 7.55 & 0.32 \\ \mr
Pristine & 2.453 & 2 & 0.00 & $P6/mmm$ (191) & $D_{6h}$ & 7.59 & 0.00 \\
\br
\end{tabular}\\
\raggedright
$^{a}$Calculations performed using the HSE06 hybrid functional.\\
$^{*}$The calculated residual band-gaps for the SWD 4$\times$4 and SWD 5$\times$5 superlattices are 0.0002 eV and 0.0007 eV, respectively.
\end{table*}
%=================================================================

\begin{figure}[ht!]
\centering
\includegraphics[width=0.7\linewidth, trim=1.6cm 0cm 0cm 0cm, clip]{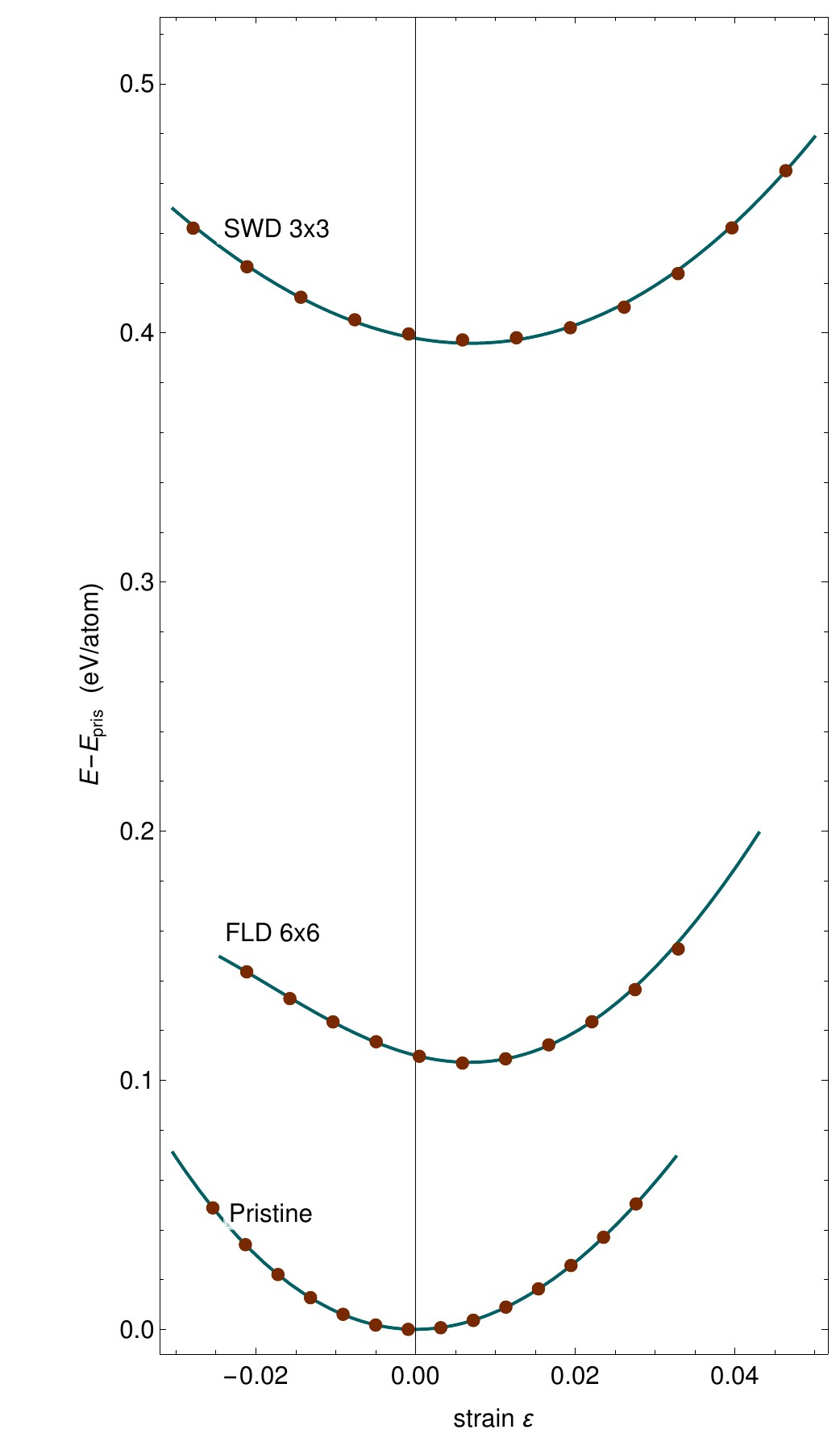}
\caption[Mechanical response of FLD and SWD superlattices]{Mechanical response of FLD 6$\times$6 and SWD 3$\times$3 superlattices. The figure shows the total energy per atom relative to pristine graphene, $E - E_{\mathrm{pris}}$, as a function of the in-plane strain $\varepsilon$ for pristine graphene.\label{fig:eos-prisSWD3x3FLD6x}}
\end{figure}

The cohesive energy per atom increases with increasing supercell size, that is, with increasing defect spacing, and gradually approaches the pristine graphene value. This trend indicates increasing energetic favorability relative to the more defect-dense superlattices. Fig.~\ref{fig:eos-prisSWD3x3FLD6x} shows the total energy per atom relative to pristine graphene, $E - E_{\mathrm{pris}}$, as a function of the applied in-plane strain $\varepsilon$\footnote{$\varepsilon$ is the dimensionless linear in-plane strain defined as $(a - a_{\mathrm{op}})/a_{\mathrm{op}}$.}. The curves are shown for pristine graphene, the SWD–3$\times$3, and the FLD–6$\times$6. All three systems exhibit a nearly parabolic dependence around their respective minima, consistent with elastic behavior in the small-strain regime. 

The energy difference relative to pristine graphene changes sign at linear strains of approximately 0.008 for FLD--6$\times$6 and 0.007 for SWD--3$\times$3. In addition, the minimum energy per atom of FLD--6$\times$6 is about 0.1~eV/atom higher than that of pristine graphene, whereas SWD--3$\times$3 is about 0.4~eV/atom higher. Taken together, these results indicate that the FLD--6$\times$6 superlattice is energetically more stable than the SWD--3$\times$3 superlattice.
However, this comparison involves different supercell sizes. For a direct comparison at the same supercell size, the SWD structures are slightly more stable than the FLD structures. In particular, for the 5$\times$5 and 6$\times$6 supercells, the cohesive energy per atom of SWD is higher by approximately 0.05 and 0.01~eV/atom, respectively, than that of the corresponding FLD superlattices.

It is worth noting that the FLD exhibits a significantly higher defect atomic fraction than the SWD at comparable supercell sizes (Defect A.F. (\%); Tab.~\ref{tab:table-general}), where the defect atomic fraction is defined as the percentage of atoms in the supercell that belong to the reconstructed defect core. This occurs because the FLD reconstruction involves 24 atoms, whereas the SWD involves only 2. Interestingly, despite the substantially larger defect atomic fraction of the FLD, the difference in cohesive energy per atom remains relatively small for the same supercell sizes. This indicates that the energetic cost per atom associated with the FLD reconstruction does not scale proportionally with the number of reconstructed atoms.

%===================================================
%===================================================

\subsection{Electronic structure of the superlattices}\label{subsec:electronic}

Here, we characterize the electronic properties of the defective graphene superlattices to determine how the presence and periodic arrangement of topological defects FLD and SWD alter the band structure and influence the possible opening of an electronic band-gap.

\subsubsection{SWD}\mbox{}\\
After structural relaxation, we analyze the electronic structure using the density of states (DOS) to study the band-gaps of the defective superlattices. As shown in Tab.~\ref{tab:table-general}, the SWD superlattices follow the superlattice selection rule, whereby a band-gap is expected only for periodicities that are multiples of three. In these cases, the Dirac points of pristine graphene fold to the same point in the supercell Brillouin zone.
Consistent with this rule, only SWD--3$\times$3 and SWD--6$\times$6 exhibit finite gaps, while the remaining periodicities remain effectively gapless. The SWD--3$\times$3 superlattice displays a clear gap of $E_g = 0.30$~eV. In contrast, the gap for SWD--6$\times$6 is nearly vanishing, with $E_g = 1$~meV (below the DOS energy-grid spacing of 7~meV). This near-gap closure is expected because decreasing the defect density drives the electronic structure toward that of pristine, gapless graphene, consistent with the low defect area fraction of SWD--6$\times$6 (2.78\%; Tab.~\ref{tab:table-general}). To validate these values, we also perform hybrid-functional (HSE06) calculations. The overall trend is preserved: SWD--$3\times3$ remains clearly gapped (0.62~eV), whereas SWD--6$\times$6 (0.06~eV) exhibits only a small gap close to the semimetallic limit.%Tab.~\ref{tab:table-general}.

%===================================================
\begin{figure*}[ht]
\centering
\includegraphics[width=0.9\linewidth]{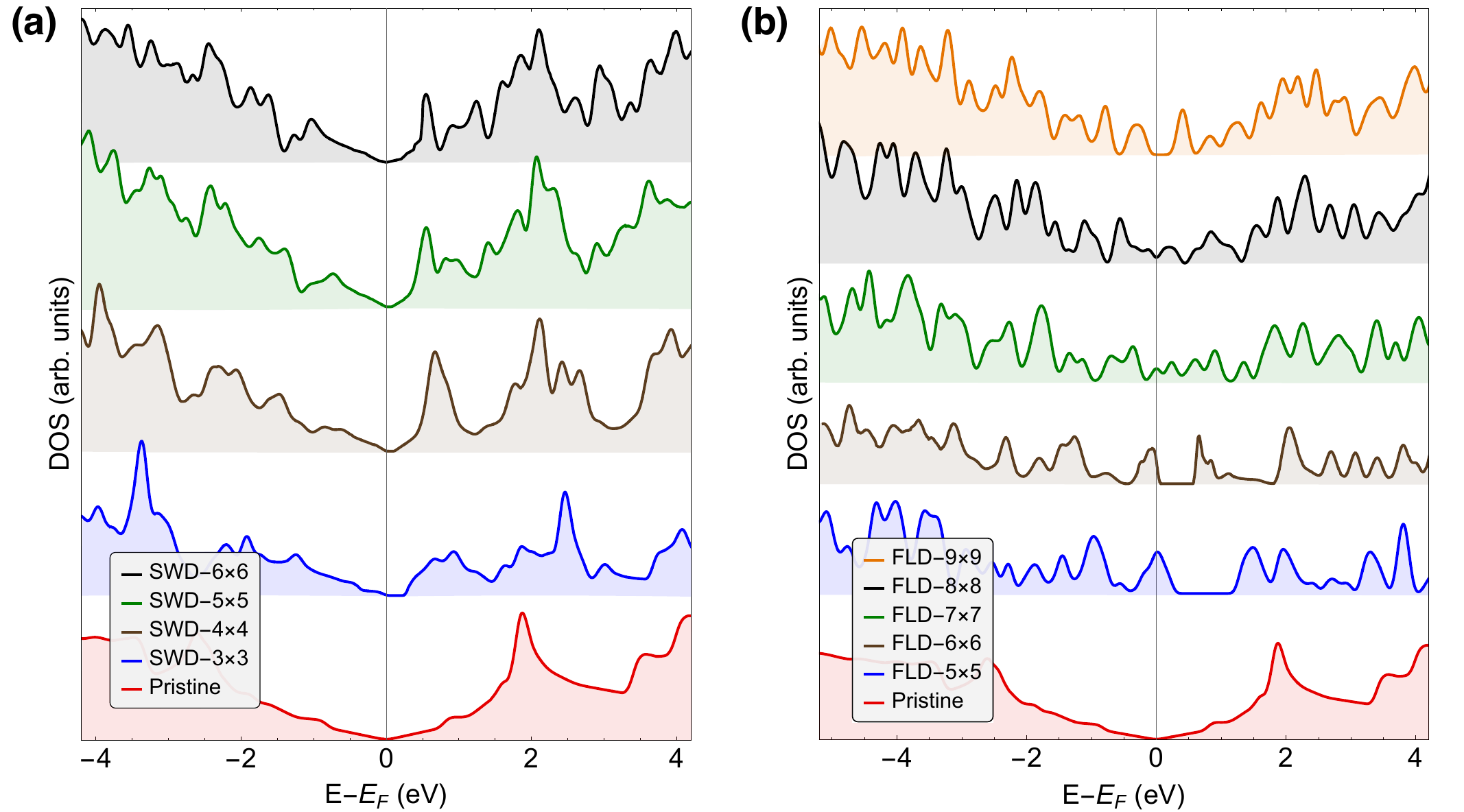}
\caption{\label{fig:dossum_all} 
$r^2$SCAN + rVV10 computed density of states (DOS) for (a) SWD and (b) FLD graphene superlattices, shown together with pristine graphene (red, bottom) for reference. The DOS curves are vertically offset for clarity. The energy axis is referenced to the Fermi level.}
\end{figure*}
%===================================================

In Fig.~\ref{fig:dossum_all}a, we present the DOS for the SWD superlattices together with pristine graphene for reference. The pristine DOS exhibits the characteristic linear suppression at the Fermi level associated with the Dirac point. While the DOS for SWD--4$\times$4 and SWD--5$\times$5 appears to show a depletion of states near the Fermi level, this feature does not correspond to a true band-gap. In defective superlattices, the Dirac crossing may be displaced from the high-symmetry lines of the reduced Brillouin zone, such that a conventional $k$-path does not pass through the actual band-touching point~\cite{shirodkar2012electronic}. To resolve this issue, we perform a refined reciprocal-space analysis by sweeping the dispersion in the vicinity of the potential Dirac point and iteratively increasing the sampling density around the relevant $K$-derived point with the smallest band-gap ($K^\ast$). This analysis confirms that SWD--4$\times$4 and SWD--5$\times$5 are gapless (see Supporting Information, Figs.~\ref{fig:SI-k_4x4}--\ref{fig:SI-k_5x5}).

\subsubsection{FLD}\mbox{}\\

Like the SWD superlattices, FLD superlattices exhibit band-gap opening only for selected periodicities. In contrast to the SWD series, however, the FLD structures display larger band-gaps and clearer gap closures at high-symmetry points. As shown in Tab.~\ref{tab:table-general}, only the FLD--6$\times$6 and FLD--9$\times$9 superlattices exhibit a band-gap, whereas the others remain effectively gapless. The calculated band-gaps are 0.62~eV for FLD--6$\times$6 and 0.32~eV for FLD--9$\times$9. This behavior is consistent with the corresponding DOS shown in Fig.~\ref{fig:dossum_all}b, which highlights the variation in electronic structure across the FLD series.

\begin{figure*}[ht]
\centering
\includegraphics[width=0.9\linewidth]{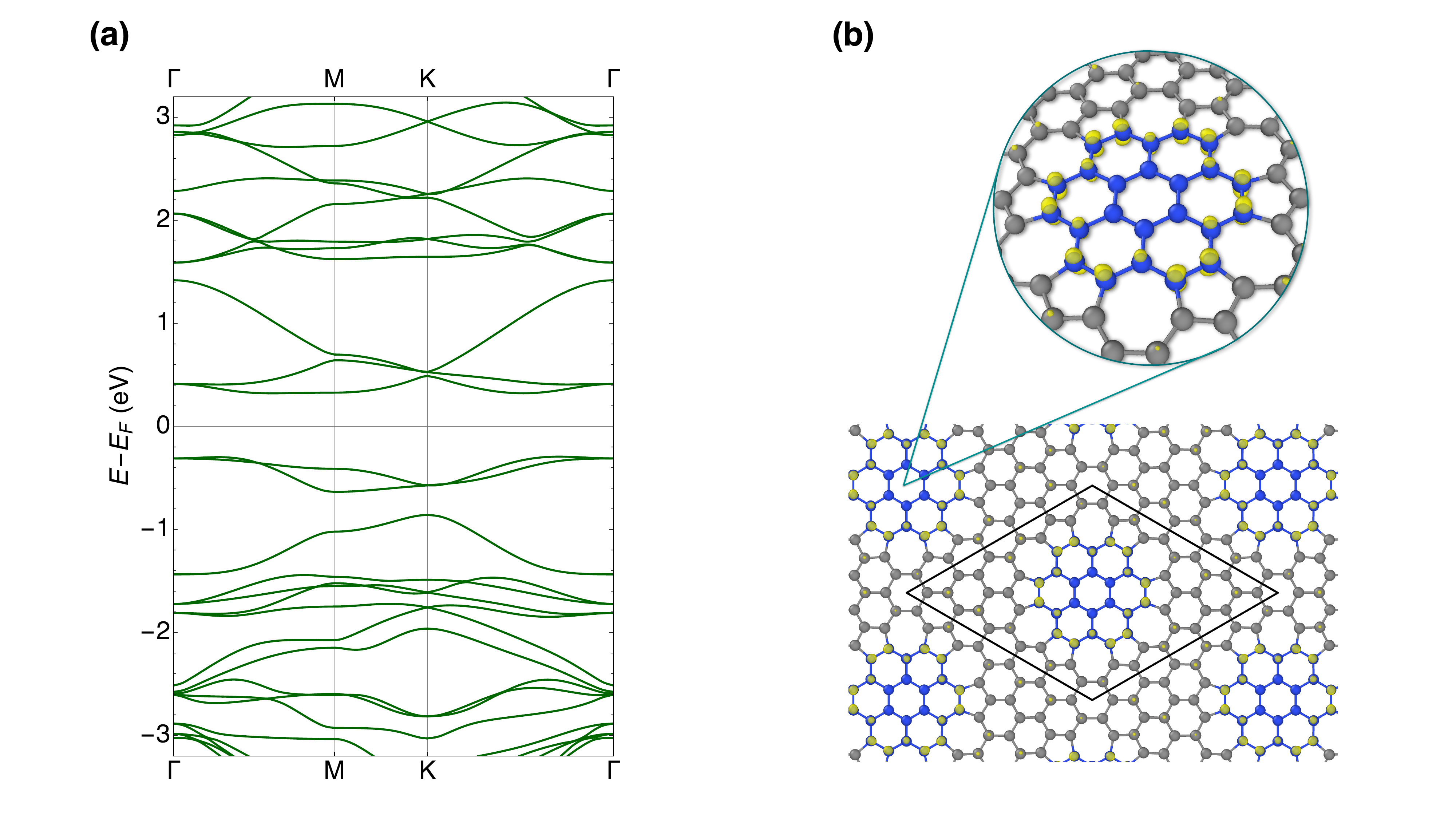}
\caption[]{Computed $r^2$SCAN+rVV10 results for the FLD--6$\times$6 superlattice. (a) Electronic band structure. (b) Top view of the relaxed structure with the contribution of the C $2p_z$ orbitals highlighted. The carbon atoms forming the FLD are shown in blue to facilitate the visualization of the flower region.}
\label{fig:fld6x6-vesta}
\end{figure*}

The electronic band structure of the FLD--6$\times$6 superlattice is shown in Fig.~\ref{fig:fld6x6-vesta}a. A direct band-gap is observed at the $\Gamma$ point in the reduced Brillouin zone, where the valence and conduction bands are separated by 0.62~eV. To further assess the accuracy of the predicted gap, the FLD--6$\times$6 system was recalculated using the HSE06 hybrid functional, which is known to provide improved band-gap estimates~\cite{heyd2003hybrid,HERNANDEZHARO2019225}. This calculation yielded a larger band-gap of 0.85~eV. 

Utilizing DFTB+, we performed calculations for the superlattices listed in Tab.~\ref{tab:table-lattice-fld-dftb}. Overall, the DFTB+ results show reasonable qualitative agreement with the DFT ($r^2$SCAN) predictions. In particular, both methods consistently reproduce the same band-gap-opening behavior for the FLD superlattices considered: FLD–5$\times$5, FLD–7$\times$7, and FLD–8$\times$8 remain gapless, whereas FLD–6$\times$6 and FLD–9$\times$9 exhibit a band-gap. Quantitatively, DFTB+ underestimates the band-gap relative to DFT by about 26\% for FLD–6$\times$6 (0.46 vs 0.62 eV) and about 22\% for FLD–9$\times$9 (0.25 vs 0.32 eV), indicating that DFTB+ should be regarded as a semi-quantitative approach for band-gap values. Nevertheless, the agreement in the gap-opening pattern supports the use of DFTB+ to test the band-gap-opening rule in larger FLD superlattices beyond the computational limitations of the r$^2$SCAN. On this basis, larger superlattices were explored using DFTB+, with the computed values summarized in Tab.~\ref{tab:table-lattice-fld-dftb}. Importantly, the DFTB+ results also support the expected superlattice selection rule, with the 12$\times$12 and 15$\times$15 periodicities exhibiting band-gaps of 0.16 and 0.11 eV, respectively. In both DFT and DFTB+, as the superlattice size increases and the defect atomic fraction decreases, the electronic structure progressively approaches that of pristine graphene, leading to smaller band-gaps that may eventually close in the dilute-defect limit.

Furthermore, in order to further understand the microscopic origin of the gap opening, we analyze the real-space electronic structure of the FLD superlattices. The bottom panel of Fig.~\ref{fig:fld6x6-vesta}b shows the periodic arrangement of flower-like defects within the graphene sheet, with the defect cores highlighted in blue to distinguish them from the surrounding hexagonal lattice. The white rhombus outlines the supercell used in the calculations, and the magnified inset emphasizes the reconstructed region of the defect, where pentagon–heptagon rings replace the pristine hexagonal network. The isosurfaces displayed above the carbon atoms correspond to the electronic charge density near the Fermi level (Fig.~\ref{fig:fld6x6-vesta}b). These isosurfaces reveal an enhanced electronic density around the reconstructed defect core, indicating that the FLD locally perturbs the $\pi$ network and modifies the electronic states in its vicinity.

\begin{figure*}[ht]
\centering
\includegraphics[width=0.9\linewidth]{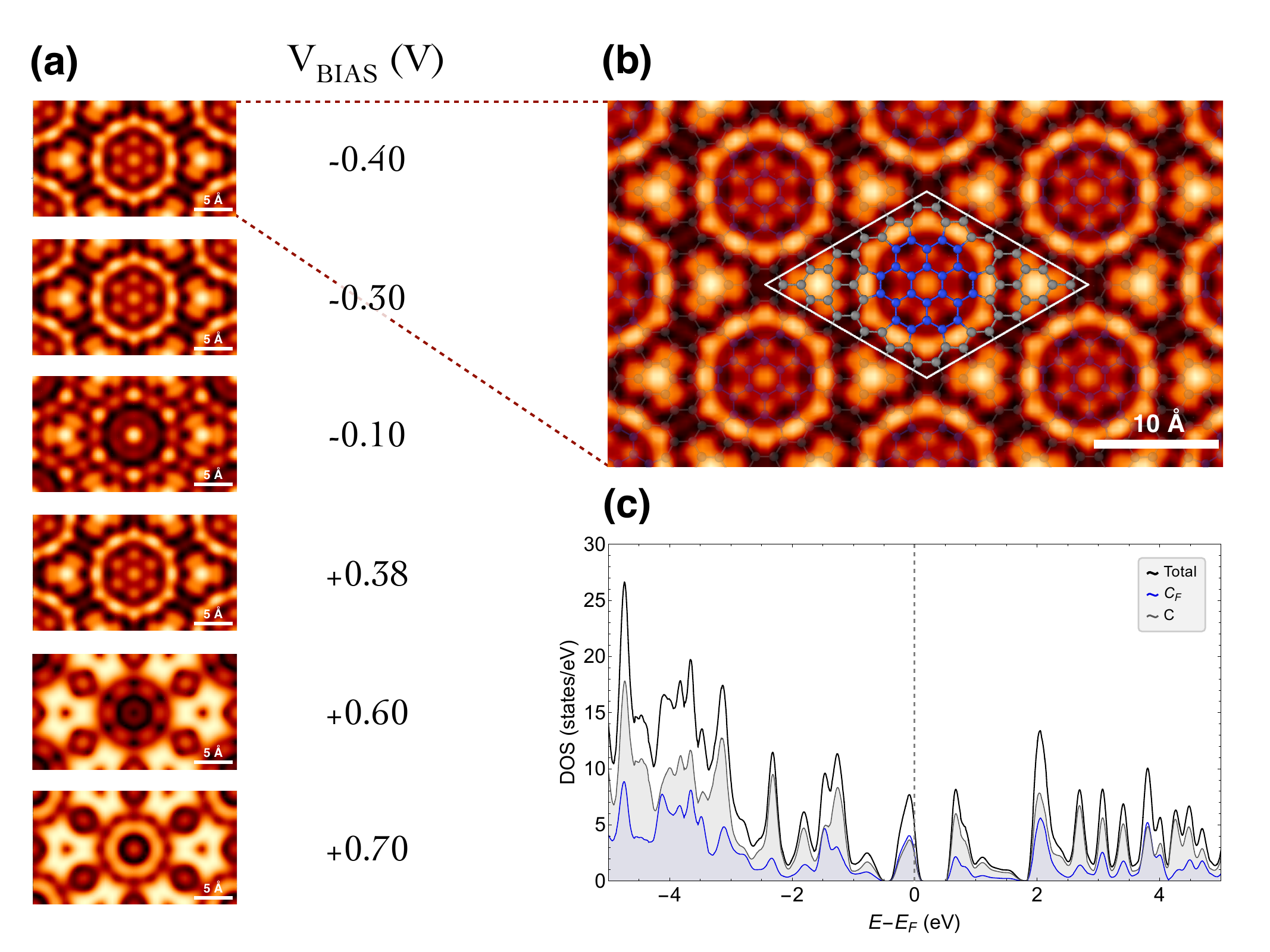}
\caption{\label{fig:stm-6x6} 
Computed STM for the FLD–6$\times$6 superlattice. 
(a) STM images for different $V_\text{bias}$ values. 
(b) Larger view of the hexagonal FLD–6$\times$6 lattice at $V_\text{bias}=-0.4$~V.
The white line denotes the unit cell of the displayed structure. 
(c) Projected DOS resolved on carbon atoms in the defective region (C$_F$, blue) and in the non-defective region (C, gray), together with the total DOS (black).}
\end{figure*}

\begin{figure*}[htbp]
\centering
\includegraphics[width=0.9\linewidth]{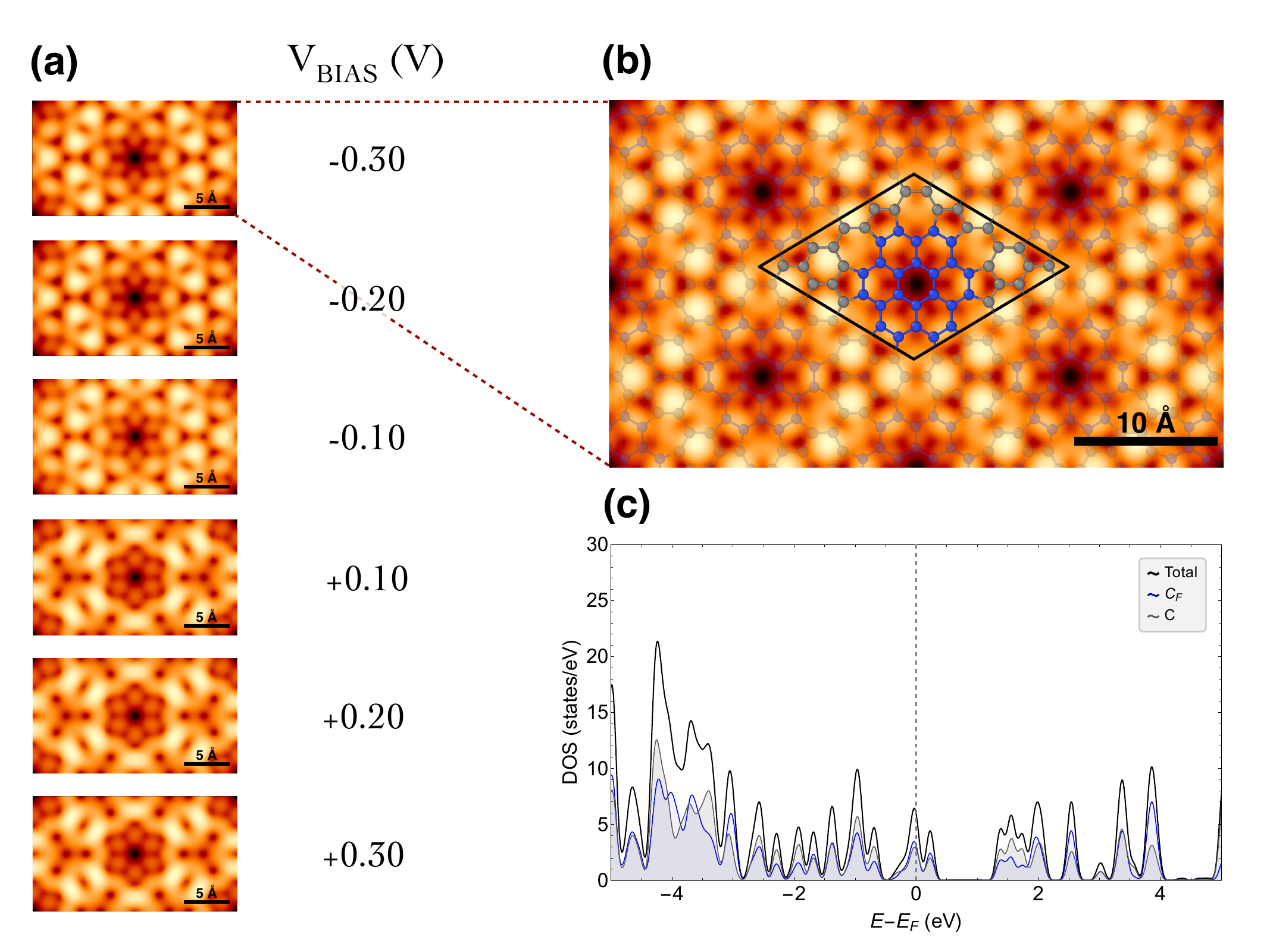}
\caption{\label{fig:stm-5x5} 
Computed STM for the FLD–5$\times$5 superlattice. 
(a) STM images for a series of bias voltages $V_\text{bias}$. 
(b) Larger view of the hexagonal FLD–5$\times$5 lattice at $V_\text{bias}=-0.3$~V.
The black line denotes the unit cell of the displayed structure. 
(c) Projected DOS resolved on carbon atoms in the defect region (C$_F$, blue) and in the non-defective region (C, gray), together with the total DOS (black).}
\end{figure*}

To connect these electronic signatures to experimentally accessible contrast, we compute simulated STM images at several bias voltages~\cite{bskan, blanco2004stm}. Fig.~\ref{fig:stm-5x5}a and Fig.~\ref{fig:stm-6x6}a show the simulated STM contrast for the FLD--5$\times$5 and FLD--6$\times$6 superlattices, respectively, for a series of $V_{\mathrm{bias}}$ values. Fig.~\ref{fig:stm-5x5}b and Fig.~\ref{fig:stm-6x6}b provide a larger field of view of the corresponding hexagonal superlattices, where the outlined polygon denotes the unit cell used in the simulation. Fig.~\ref{fig:stm-5x5}c and Fig.~\ref{fig:stm-6x6}c report the projected DOS resolved on carbon atoms in the defective region ($C_F$) and in the non-defective region ($C$), together with the total DOS, with all energies referenced to the Fermi level. 

%===================================================
%===================================================

%=====================================================
%=====================================================

A direct comparison between Fig.~\ref{fig:stm-6x6} and the real-space electronic features in Fig.~\ref{fig:fld6x6-vesta}b indicates that the enhanced apparent height over the defect region is governed by localized $\pi$ states with dominant $2p_z$ character. In the relevant bias window, these $p_z$-derived states are strongly concentrated on the carbon atoms forming the FLD boundary, and their spatial orientation deviates from the largely perpendicular $p_z$ alignment in pristine graphene. Consequently, defect-localized $2p_z$ states provide the dominant contribution to the STM signal in the vicinity of the flower-like defect. Importantly, 
these modified defect-localized orbitals provide the short-range perturbation that can couple the Dirac cones when the superlattice periodicity folds them onto the same momentum, thereby enabling gap opening.
%these modified orbitals around the defects are responsible for coupling the Dirac cones and inducing a gap. % too strong

\begin{table}[htbp]
\centering
\caption{\label{tab:table-lattice-fld-dftb} 
Computed lattice parameters and energies of FLD superlattices using the DFTB+ code. 
$a_{\mathrm{op}}$ is the optimum lattice constant, $N_{\mathrm{atom}}$ is the number of atoms, 
$E_{c}$ is the cohesive energy per atom, and $E_{g}$ is the electronic band-gap.}
\begin{tabular}{@{}lcccc}
\br
System & $a_{\mathrm{op}}$ (\AA) & $N_{\mathrm{atom}}$ & $E_{c}$ (eV) & $E_{g}$ (eV) \\
\mr
Pristine   & 2.451  &   2 & 8.684 & 0.00 \\
FLD–5$\times$5& 12.358 &  50 & 8.524 & 0.00 \\
FLD–6$\times$6 & 14.782 &  72 & 8.582 & 0.46 \\
FLD–7$\times$7& 17.224 &  98 & 8.611 & 0.00 \\
FLD–8$\times$8& 19.664 & 128 & 8.629 & 0.00 \\
FLD–9$\times$9& 22.106 & 162 & 8.640 & 0.25 \\
FLD–10$\times$10& 24.550 & 200 & 8.649 & 0.00 \\
FLD–11$\times$11 & 26.995 & 242 & 8.655 & 0.00 \\
FLD–12$\times$12 & 29.441 & 288 & 8.660 & 0.16 \\
FLD–15$\times$15 & 36.781 & 450 & 8.668 & 0.11 \\
\br
\end{tabular}
\end{table}

\subsection{Symmetries}\label{subsec:sym}

The corresponding space groups, obtained from $r^2$SCAN+rVV10  calculations, are summarized in Tab.~\ref{tab:table-general}. Each space group is listed by its Hermann–Mauguin symbol (e.g., $P6/mmm$, $P6mm$, $Cmmm$) along with its international number from the International Tables for Crystallography. For instance, $P6/mmm$ (No. 191) denotes a hexagonal lattice with mirror planes and sixfold rotational symmetry.

In the SWD superlattices, the defects reduce the sixfold rotational symmetry of graphene to a twofold rotation symmetry, leading to the space group $Cmmm$. FLD lattices, on the other hand, preserve the sixfold rotation symmetry and therefore have space groups $P6/mmm$ or $P6mm$ (the difference between these groups arises from out-of-plane distortions after lattice relaxation). Importantly, both SWD and FLD lattices preserve $C_2$ rotation symmetry, which protects isolated Dirac cones from gapping in graphene~\cite{RMP_2009}. As we discuss below, the symmetries of the superlattice are intimately related to the presence or absence of a band-gap. Indeed, symmetry considerations provide a useful guide for identifying which structures are expected to develop a gap in general.

\subsection{ Mechanism for gap opening}\label{subsec:mechanism}

In previous works, it was argued that periodic adatom decoration in graphene can create a confinement potential for charge carriers in pristine graphene regions, resulting in the opening of a band-gap~\cite{Balog2010}. However,
Dvorak \textit{et al.}~\cite{Dvorak2013} proposed a perturbative tight-binding model that suggests band-gap opening in graphene with arbitrary Bravais periodic patterning is not solely due to this confining effect, but rather appears only for certain superlattice periodicities. Our results, which concern graphene patterned with periodic arrays of defects, agree with those of Dvorak \textit{et al.} \cite{Dvorak2013} in that only certain superlattices lead to gap opening.

By applying a tight-binding model to graphene with grain boundary loops, we derived a simple rule for band-gap opening in Eq.~\ref{eq:ruleFLD}: superlattices with lattice vectors that are multiples of three exhibit a finite band-gap. The DFT results confirm band-gap opening for the 6$\times$6 and 9$\times$9 FLD superlattices, and complementary DFTB+ results for the 12$\times$12 and 15$\times$15 superlattices also support this rule. Nevertheless, as the size of the supercell increases, the electronic structure progressively resembles that of pristine graphene, leading to smaller gaps that may eventually close.

The Dirac points in pristine graphene are protected by a combination of time-reversal symmetry, inversion symmetry, and translation symmetry~\cite{RMP_2009}. 
Breaking any of these symmetries can lead to a gap opening at the Dirac points. Without heightened electronic correlations, as in twisted bilayer graphene~\cite{kuiri2022time} or rhombohedral pentalayer graphene~\cite{han2024rhombohedral}, breaking time-reversal symmetry is difficult. On the other hand, inversion and translation symmetry may be broken simply by modifying the structure of the lattice. 

In particular, defect-patterned graphene superlattices can open a gap in two ways~\cite{park2015band}. First, the defects can themselves induce an asymmetry between the $A$ and $B$ sublattices of the hexagonal lattice, thus breaking inversion symmetry. As a result, the independent Dirac cones at $K$ and $K'$ are no longer protected and can be gapped. If this were the case in FLD and SWD lattices, however, a gap should open for any periodicity of the defective lattice. Our findings indicate that this is not the case, since only some periodicities lead to gap opening. Indeed, both Stone-Wales and Flower-like defects preserve $C_2$ symmetry, as confirmed by our symmetry analysis in Sec.~\ref{subsec:sym}, which in two dimensions is the same as inversion symmetry.
Consequently, the gaps observed in FLD and SWD lattices must arise from another mechanism.

The second way in which graphene superlattices can open a gap is by breaking translation symmetry. Any superlattice has a larger unit cell than pristine graphene, leading to a reduced translation symmetry. As the unit cell is enlarged, the Brillouin zone is folded. In this process, two distinct Dirac points at $K$ and $K'$ can both fold to the same crystal momentum. If this is the case, short-wavelength perturbations coupling the $K$ and $K'$ Dirac cones can then open a gap, as depicted in Fig.~\ref{fig:bzfold}. In the literature, this is sometimes referred to as a Kekulé mass in graphene~\cite{chamon2012_mass}.

Because the FLD and SWD lattices preserve $C_2$ symmetry, this latter mechanism is the only possibility for opening a gap. All defective superlattices that we consider break the translation symmetry of pristine graphene. However, as we have shown in Sec.~\ref{sec:methods}, only supercells of size $3N\times 3N$ fold $K$ and $K'$ to the same momentum ($\Gamma$) in the new Brillouin zone. As a result, these are the only structures for which a gap can open. Altered bonds near defect cores, as shown in Fig.~\ref{fig:fld6x6-vesta}(b), then couple the two Dirac cones and open a gap.

With an understanding of the pattern of symmetry-breaking required to open a gap in graphene, we can also predict gap opening in other defect lattices. In particular, if the defect lattices preserve $C_2$ symmetry, we have shown that only certain supercells may open a gap. On the other hand, if the lattices themselves break $C_2$ symmetry, a gap can form for any supercell size. Such guiding principles are helpful in identifying promising periodically reconstructed lattices for band-gap engineering.

\section{Conclusions}\label{sec:conclusions}
In this work, we investigated the structural and electronic properties of graphene superlattices patterned with periodic Stone–Wales (SWD) and flower-like (FLD) topological defects. Using DFT calculations supported by tight-binding analysis, we systematically examined how defect type and periodicity influence lattice stability, Brillouin zone folding, and band-gap formation. We identified the symmetry conditions under which a band-gap emerges and clarified the role of defect-induced modifications in the Dirac cones. Our results establish a predictive framework linking superlattice periodicity, symmetry and controllable band-gap engineering in graphene. The main conclusions are as follows:

\begin{itemize}

\item Among the investigated SWD periodicities, only the SWD--3$\times$3 and SWD--6$\times$6 superlattices exhibit finite band-gaps. The SWD--3$\times$3 structure displays a clear gap of $E_g = 0.30$~eV, whereas the SWD--6$\times$6 gap is nearly vanishing ($E_g = 0.001$~eV), consistent with its larger defect separation. Hybrid-functional (HSE06) calculations confirm this trend, yielding enhanced but qualitatively consistent band-gaps for both periodicities. Apparent gap-like features in SWD--4$\times$4 and SWD--5$\times$5 originate from $k$-space sampling effects. A refined reciprocal-space analysis confirms that these structures remain gapless, in agreement with the literature.

\item Finite band-gaps are obtained for FLD--6$\times$6 and FLD--9$\times$9, with $E_g = 0.62$~eV and $E_g = 0.32$~eV, respectively, while the remaining periodicities remain effectively gapless. Hybrid-functional (HSE06) calculations increase the predicted gap of FLD--6$\times$6 to 0.85~eV, while preserving the qualitative band structure and selection-rule behavior. Extended DFTB+ calculations confirm that the selection rule persists for larger periodicities, with band-gaps appearing for FLD--12$\times$12 and FLD--15$\times$15.

\item  Real-space charge-density analysis and simulated STM images show that defect-localized $2p_z$ states are concentrated around the reconstructed defect core and dominate the local tunneling contrast, providing an experimentally accessible signature of the defect-induced modification of the electronic structure.

\item Both SWD and FLD superlattices follow the same selection rule: a band-gap opens only for periodicities that are multiples of three, for which the Dirac points of pristine graphene fold onto the same momentum in the reduced Brillouin zone. For both defect types, increasing the separation between defects reduces the perturbation of the graphene $\pi$ network, causing the electronic structure to converge toward that of pristine graphene and the band-gap to diminish in the dilute-defect limit.

\item Both SWD and FLD superlattices reduce the translation symmetry of pristine graphene by enlarging the unit cell, while preserving specific point-group symmetries depending on the periodicity.

\item Symmetry analysis using FINDSYM shows that the relaxed SWD and FLD superlattices retain $C_2$ rotation symmetry, which in two dimensions is equivalent to inversion symmetry. For the defect superlattices studied here, band-gap opening is therefore not driven by inversion-symmetry breaking (i.e., a Semenoff mass~\cite{semenoff1984Dec}), but by the reduction of translation symmetry, which folds the $K$ and $K'$ valleys onto the same momentum in the reduced Brillouin zone and allows intervalley coupling (i.e., a Kekul\'e mass).

\end{itemize}

We note that the present study is based on idealized,  periodic defective superlattices. In realistic samples, disorder, finite-temperature effects, substrate interactions, and strain may influence both the stability of the structures and the magnitude of the band-gap. In particular, the controlled creation of topological defects such as SWD and FLD in large-scale, periodically ordered arrays remains experimentally challenging. 

Overall, our results demonstrate that the electronic band-gap in graphene can be predictively controlled through the deliberate choice of defect type, superlattice periodicity, and defect separation. It is therefore possible to engineer finite and tunable band-gaps while preserving the underlying graphene framework. This symmetry-guided defect patterning provides a potential route toward the design of graphene-based nanoelectronic devices.

\section*{\label{acnow} Acknowledgments}
We would like to express our gratitude to Prof. Rob Leigh, Prof. Alicja Mikołajczyk, and Jorge Vega for their valuable comments and insightful discussions. Additionally, we extend our acknowledgment for the computing time provided by the TASK supercomputer center in Gdańsk, Poland.
% Acknowledging Australian resources
D.~N.~Garzon acknowledges support from the AAUW Doctoral Fellowship, the ICASU Physics Fellowship from the University of Illinois.
LC, MF and CS gratefully acknowledge the financial support of the Australian Research Council (ARC FL230100176). Partial theoretical calculations in this research were undertaken with the assistance of resources from the National Computational Infrastructure (NCI), which is supported by the Australian Government.

\bibliographystyle{unsrt}
\bibliography{references} 
%\clearpage
%\pagestyle{empty} 

% ===================== SI format =====================
\clearpage
\onecolumn
\markboth{SUPPORTING INFORMATION}{SUPPORTING INFORMATION}
\setcounter{section}{0}
\setcounter{equation}{0}
\setcounter{figure}{0}
\setcounter{table}{0}
\renewcommand{\thesection}{S\arabic{section}}
\renewcommand{\theequation}{S\arabic{equation}}
\renewcommand{\thefigure}{S\arabic{figure}}
\renewcommand{\thetable}{S\arabic{table}}
%\nolinenumbers

\section*{\Large{Supporting Information} \\
\vspace{0.2cm}
\large{Symmetry Guided Band-Gap Opening via Periodic Topological Defects in Graphene}}
\vspace{-0.2cm}
\normalsize{
D. N. Garzon$^{1,2,3 \ast, \ddag}$, 
Leonel Cabrera-Loor$^{1,4, \ast, \S}$, 
Jacopo Gliozzi$^{3}$, 
Marco Fronzi$^{4}$, 
Catherine Stampfl$^{4}$ and
Henry P. Pinto$^{1}$
}
\address{$^1$ CompNano Group, School of Physical Sciences and Nanotechnology, Yachay Tech University, 100119-Urcuqui, Ecuador}
\address{$^2$ Illinois Center for Advanced Studies of the Universe, University of Illinois Urbana-Champaign, Urbana, Illinois 61801, USA}
\address{$^3$The Grainger College of Engineering,
Department of Physics, University of Illinois Urbana-Champaign, Urbana, Illinois 61801, USA}
\address{$^4$ School of Physics, The University of Sydney, Camperdown, NSW 2006, Australia}
\begin{indented}
\item[]$^\ast$ These authors contributed equally to this work.
\item[]$^\ddag$ Present address: Department of Physics, University of Illinois at Urbana-Champaign, Urbana, IL 61801, USA.
\item[]$\S$ Present address: School of Physics, The University of Sydney, Camperdown, NSW 2006, Australia.
\end{indented}

\section*{Convergence tests for \texttt{ENCUT} and \texttt{KPOINTS} selection}
\begin{figure}[ht]
\centering
\includegraphics[width=0.6\linewidth]
{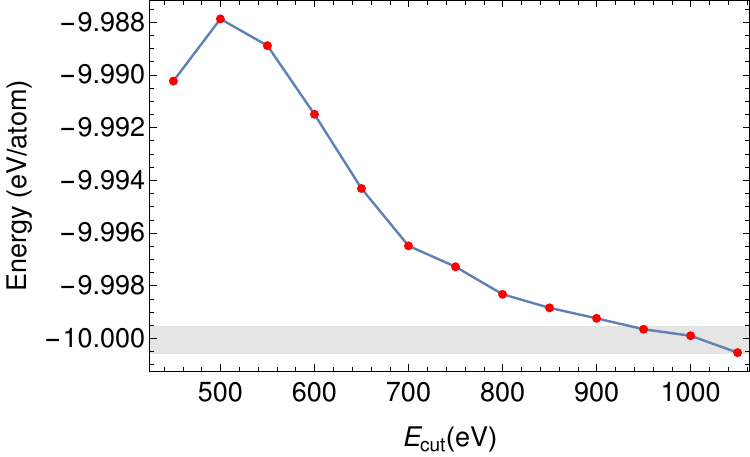}
\caption{Convergence of the total energy per atom as a function of plane-wave kinetic-energy cutoff ($E_{\mathrm{cut}}$). The grey band indicates the 1~meV~atom$^{-1}$ convergence criterion used to select the final cutoff. The line connecting data points is a guide to the eye.}
\label{fig:cutoff}
\end{figure}

\begin{figure}[ht]
\centering
\includegraphics[width=0.6\linewidth]
{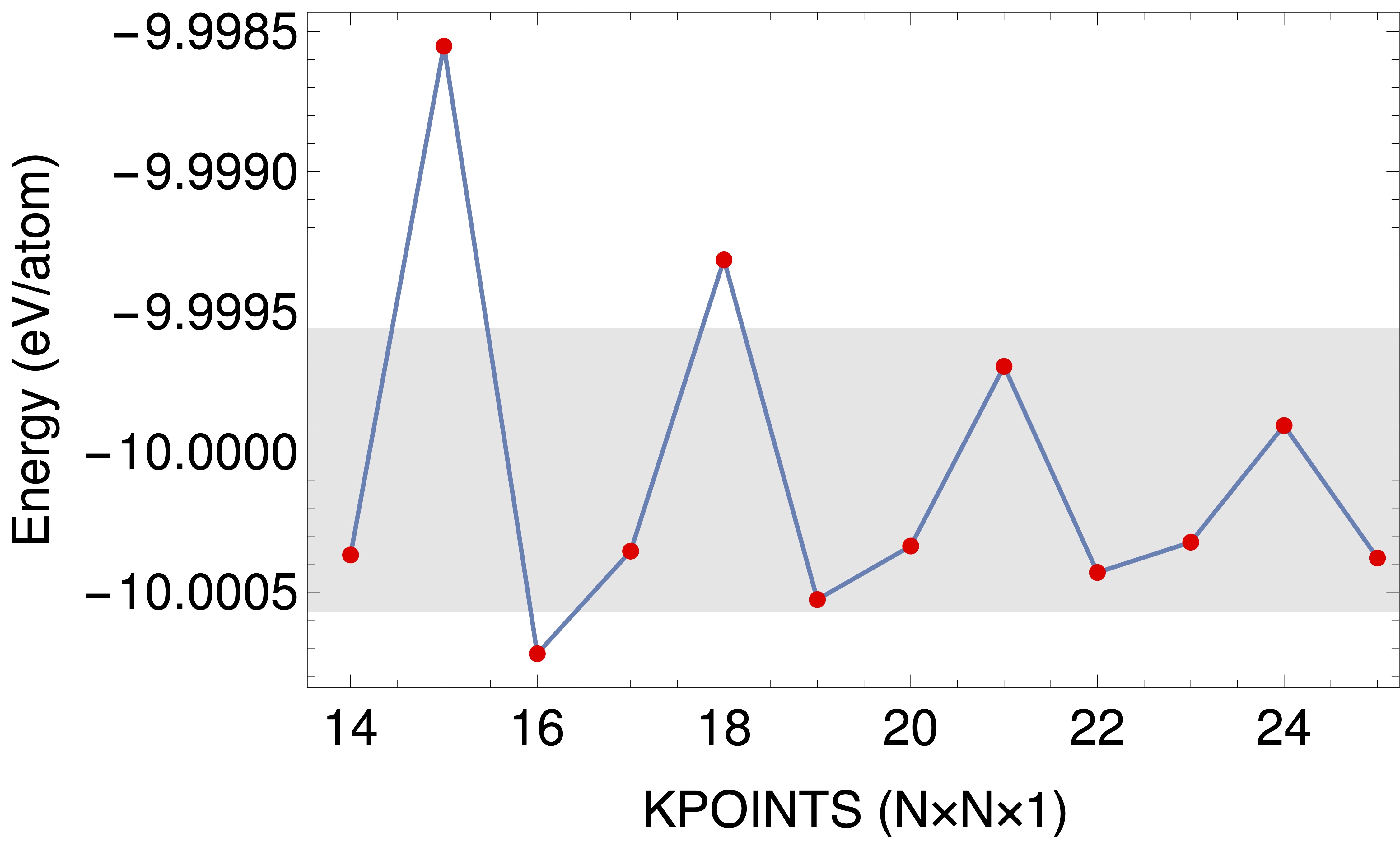}
\caption{Total-energy per atom convergence as a function of the $k$-point sampling. The grey band denotes the 1~meV convergence criterion. For pristine graphene, a 21$\times$21$\times$1 mesh was adopted, as it includes the Dirac point. This mesh corresponds to a $k$-point spacing of $0.022\times 2\pi$~\AA$^{-1}$, which was used for all other superlattices.}
\label{fig:kpts}
\end{figure}

\section*{Refined Reciprocal-Space for 4$\times$4 and 5$\times$5 SWD superlattices}

The DOS shown in the main paper (Fig.~\ref{fig:dossum_all}) for SWD--$4\times4$ and SWD--5$\times$5 exhibit a depletion of states near the Fermi level, which appears inconsistent with the expected $3N\times3N$ selection rule. We initially suspected that this feature arose from limited energy resolution or insufficient $k$-point sampling. However, increasing the energy resolution of both the DOS and the band-structure analysis down to 1~meV still did not recover the expected band touching. We therefore attribute this apparent pseudogap to a Dirac crossing that is displaced from the high-symmetry lines of the reduced Brillouin zone, such that a conventional $k$-path does not pass through the true band-touching point~\cite{shirodkar2012electronic}. However, adopting the alternative $k$-path suggested in Ref.~\cite{shirodkar2012electronic} did not resolve the true Dirac point in our models.

\begin{figure*}[h!]
\centering
\includegraphics[width=1\linewidth]{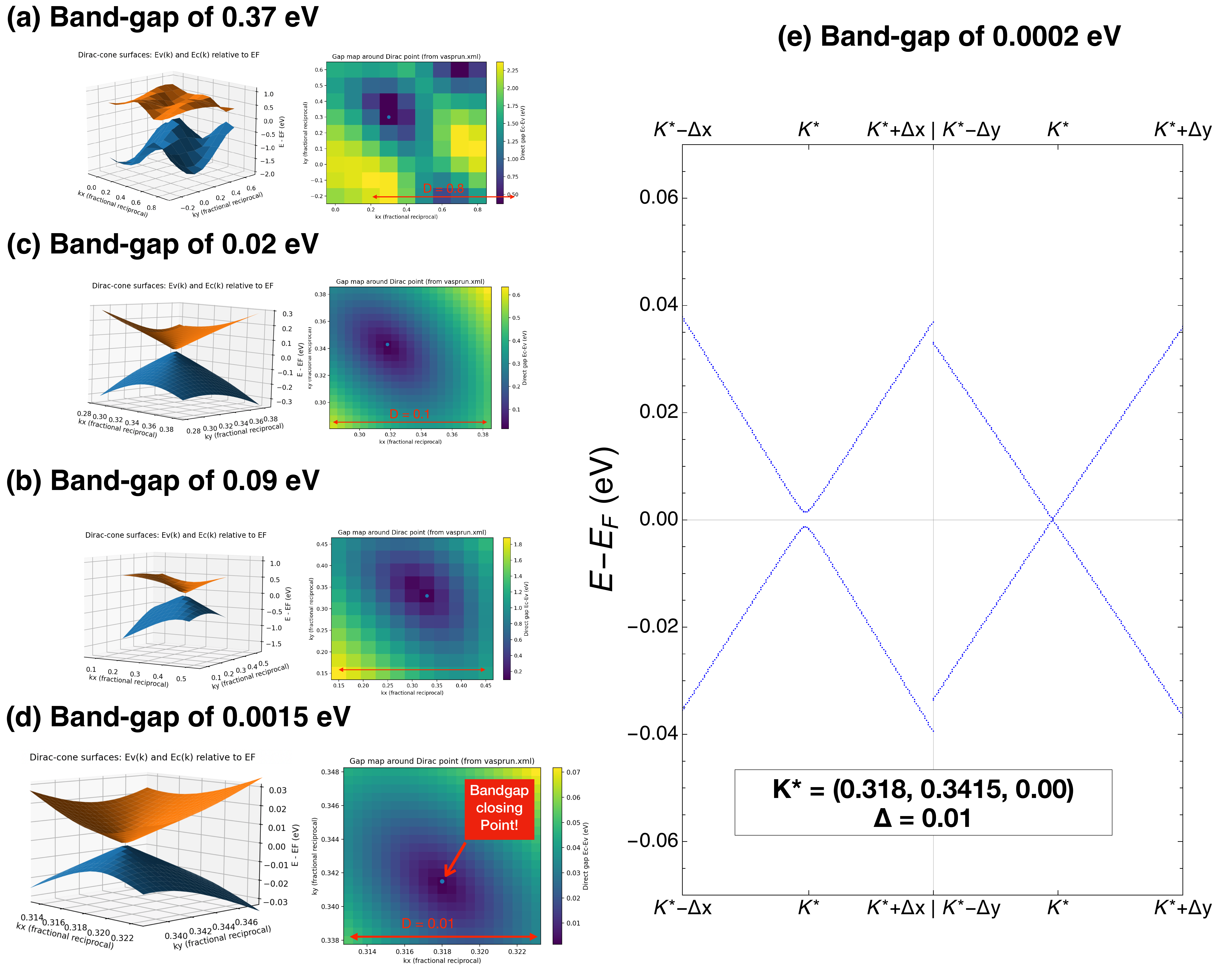}
\caption{Progressive $k$-space refinement for the SWD--4$\times$4 superlattice demonstrating band-gap closure near the Dirac point. Panels (a--d) show successive zoom-ins of the conduction- and valence-band surfaces $E_c(\mathbf{k})$ and $E_v(\mathbf{k})$ (left) together with the corresponding gap map $\Delta E(\mathbf{k}) = E_c - E_v$ (right), where the apparent gap decreases from 0.37~eV to 0.0015~eV as the sampling window is narrowed around the minimum-gap region. The refined search identifies the band-touching point $K^\ast$ (marked in panel (d)). Panel (e) reports the final high-resolution band structure computed along $K^\ast\!\pm\!\Delta k_x$ and $K^\ast\!\pm\!\Delta k_y$, using a line-mode path centered at $K^\ast=(0.318,\,0.3415,\,0.0)$ with a half-window $\Delta=0.01$, and $N=200$ $k$-points per segment, yielding a residual gap of 0.2 meV.}
\label{fig:SI-k_4x4}
\end{figure*}

\begin{figure*}[h!]
\centering
\includegraphics[width=\linewidth]{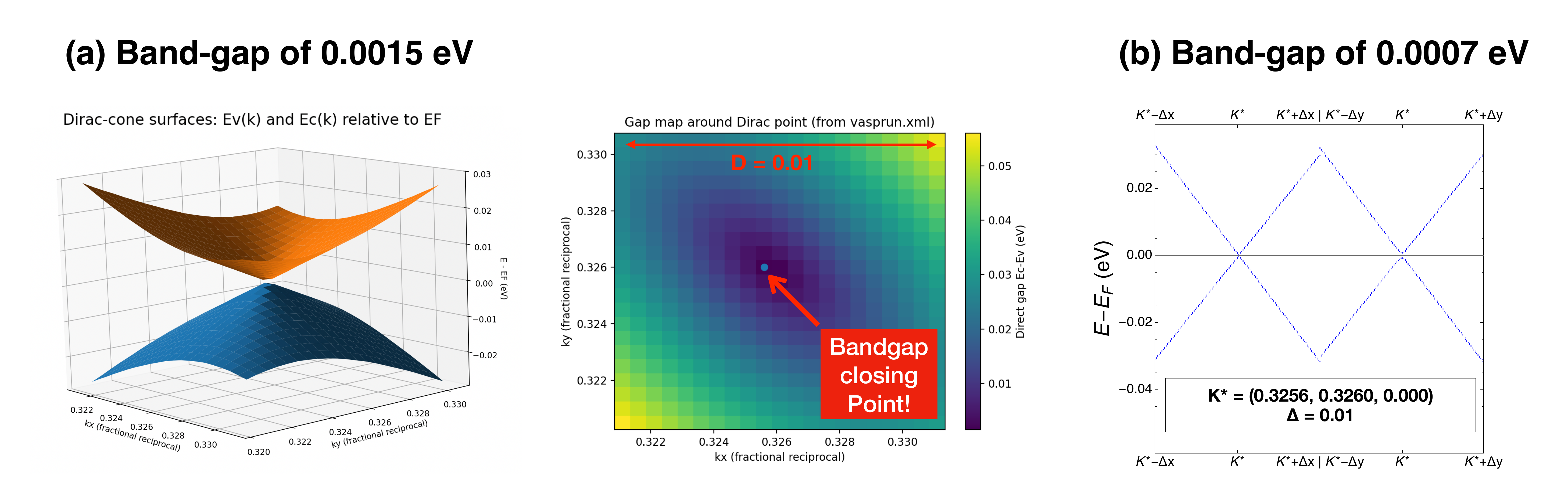}
\caption{$k$-space refined analysis for the SWD--5$\times$5 superlattice. Panel (a) shows the last refined conduction- and valence-band surfaces $E_c(\mathbf{k})$ and $E_v(\mathbf{k})$ in the vicinity of the minimum-gap point, together with the corresponding gap map $\Delta E(\mathbf{k})=E_c-E_v$, highlighting the band-touching region at $K^\ast$. Panel (b) reports the high-resolution band structure evaluated along $K^\ast\!\pm\!\Delta k_x$ and $K^\ast\!\pm\!\Delta k_y$, using a line-mode path centered at $K^\ast=(0.3256,\,0.3260,\,0.0)$ with a half-window $\Delta=0.01$, and $N=200$ $k$-points per segment, confirming band-gap closure within numerical tolerance (residual gap 0.7 meV).}
\label{fig:SI-k_5x5}
\end{figure*}

To resolve this issue, we performed a refined reciprocal-space analysis by sweeping the dispersion in the vicinity of the Dirac point and iteratively increasing the sampling density around the relevant $K$-derived point having the smallest band-gap ($K^{\ast}$). This analysis confirms that SWD--4$\times$4 and SWD--5$\times$5 are gapless, as clearly shown in Figs.~\ref{fig:SI-k_4x4} and \ref{fig:SI-k_5x5}.
%%%%%%%%%%%%
%%%%%%%%%%%%

\end{document}